\documentclass{elsart}

 \usepackage{epsfig}

\usepackage{amssymb}

\begin{document}

\begin{frontmatter}



 \corauth[cor1]{Corresponding author:
D. Popovi\'{c}, phone 1-850-644-3913, fax 1-850-644-5038}

\title{Glass Transition in a Two-Dimensional Electron System in 
Silicon}


\author{Sne\v{z}ana Bogdanovich,}
\author{Dragana  Popovi\'c}
 \ead{dragana@magnet.fsu.edu}
\address{National High Magnetic Field Laboratory, Florida State
University\\ 1800 East Paul Dirac Drive, Tallahassee, FL 32310, USA}
\begin{abstract}
Large fluctuations of conductivity with time are observed in a
low-mobility two-dimensional electron system in silicon 
at low electron densities $n_s$ and temperatures. 
A dramatic increase of the noise power ($\propto 1/f^{\alpha}$) as $n_s$ is 
reduced below a certain density $n_g$, and a sharp jump of $\alpha$
at $n_s\approx n_g$,
are attributed to the freezing of the electron glass at $n_s = n_g$.  The data
strongly suggest that glassy dynamics persists in the metallic phase.
\end{abstract}

\end{frontmatter}

\section{Introduction}

The metal-insulator transition (MIT) in two dimensions (2D) has been a subject
of intensive research in recent years~[1] but the physics behind this
phenomenon is still not understood.  It has been established, however, that it
occurs in the regime where both Coulomb (electron-electron) interactions 
and disorder are strong.  The competition between these two effects has been
suggested [2] to 
lead to glassy dynamics (electron or Coulomb glass).  Therefore, it has been 
proposed that the 2D MIT can be described alternatively as the melting of the 
Wigner glass [3], or the melting of the Coulomb glass [4].  The existence of a
large number of metastable states in a glass results in fluctuations of
conductivity $\sigma$ with time (conductivity noise) so that mesoscopic 
samples should be more suitable for studies of glassy properties.  Indeed, 
investigations of metallic spin glasses have shown [5] that mesoscopic 
measurements are necessary in order to provide detailed understanding of 
glassy ordering and dynamics.  Here we report the results of transport and 
noise studies in mesoscopic samples, which demonstrate for the first time the 
existence of a glassy phase in a 2D system in semiconductor heterostructures.

\section{Experiment}

The measurements were performed on a 1~$\mu$m long, 90~$\mu$m wide rectangular
n-channel Si metal-oxide-semiconductor field-effect transistor (MOSFET) with 
the peak mobility of only $0.06~$m$^{2}$/Vs at 4.2~K. Other sample details 
have been given elsewhere [6]. The fluctuations of current $I$ ({\it i.~e.} 
$\sigma$) as a function of time were measured using a low-noise
current preamplifier and a standard two-probe lock-in technique at 
$\sim 13$~Hz with the constant excitation voltage $V_{exc}\sim$ a few $\mu$V.
By changing the gate voltage $V_g$, the electron density $n_s$ was varied 
between $3.4\times 10^{11}$cm$^{-2}$ and $20.2\times 10^{11}$cm$^{-2}$, 
whereas temperature $T$ ranged between 0.13~K and 0.80~K.  It was 
established that the current fluctuations were not correlated with ({\it 
i.~e.} not due to) the fluctuations of temperature or applied voltage.  In
addition, another sample from the same wafer was measured in a four-terminal 
configuration at 0.25~K, and it was determined that the contact resistances and
the contact noise were negligible.  

\section{Results}

Fig. 1 (left) shows the time dependence of the relative fluctuations 
$(\sigma-\langle\sigma\rangle)/\langle\sigma\rangle$ (where 
$\langle\ldots\rangle$ denotes averaging over time) 
for a fixed $n_s$ at different $T$.
At the lowest $T=0.13$~K, fluctuations are as high as 
250~$\%$, and drop rapidly with increasing $T$. A similar
decrease in fluctuations was observed with increasing $n_s$, as discussed in 
more detail below.

The time-averaged conductivity $\langle\sigma\rangle$ was calculated for time 
intervals of several hours, for all $n_s$ and $T$.  Fig. 1 (right) depicts 
$\langle\sigma\rangle$ as a function of $T$ at selected $n_s$. 
Similar to the behavior of various high-mobility 2D systems~[1], the 
temperature coefficient of conductivity $d\langle\sigma\rangle/dT$ changes 
sign when
$\langle\sigma(n_{s}^{\ast})\rangle=0.5~e^2/h$.  Even though the corresponding
density $n_{s}^{\ast}=12.9\times10^{11}$cm$^{-2}$ is much higher due to a 
large amount of disorder in our samples, the effective Coulomb interaction is 
still comparable to that in other 2D systems ($r_s=4.6$, $r_s$ -- ratio of 
Coulomb energy to Fermi energy).  The densities $n_{s}^{\ast}$, where 
$d\langle\sigma\rangle/dT=0$, have been usually~[1] identified with the 
critical density $n_c$ for the metal-insulator transition.  However, a 
thorough analysis of $\langle\sigma (n_s,T)\rangle$ at low $n_s$ shows~[6] 
that, in our case,
$n_c=(5.0\pm 0.3)\times 10^{11}$cm$^{-2}$, which is more than a factor of two
smaller than $n_{s}^{\ast}$. The difference between $n_c$ and $n_{s}^{\ast}$ 
in our samples is attributed to a larger amount of disorder.

The root-mean-square (rms) fluctuations 
$\delta\sigma=\langle(\sigma-\langle\sigma\rangle)^2\rangle^{1/2}$ (calculated
over the frequency bandwidth from (10 hours)$^{-1}$ to (6 seconds)$^{-1}$) 
increase with $n_s$, 
ranging between $(10^{-5}-10^{-2})~e^2/h$ at all $T$.  Fig. 2 (left)
shows the relative rms, $\delta\sigma/\langle\sigma\rangle$, as a function of 
$n_s$ at three different $T$.  
While fluctuations seem to be independent of density at high $n_s$, a 
dramatic increase of $\delta\sigma/\langle\sigma\rangle$ 
is observed with decreasing $n_s$ below
$n_g = (7.5\pm 0.3) \times10^{11}$cm$^{-2}$.
Even though an increase in $T$ causes a substantial reduction of 
$\delta\sigma/\langle\sigma\rangle$, the onset of large fluctuations
does not seem to depend on temperature.

The noise was studied in more detail using normalized power 
spectra $S_I(f)=S(I,f)/I^{2}$, most of which were obtained in the 
$f=10^{-4}-10^{-1}$~Hz bandwidth.  They were found to follow the well-known
power-law frequency dependence $S_I\propto 1/f^{\alpha}$ (Hooge's law) [7].
In all measurements, the device noise was extracted from the total measured
noise by subtracting the background noise present with no current flowing, {\it
i.~e.} by setting $V_{exc}=0$.  The power spectrum of the background noise
was always several orders of magnitude smaller than sample noise and had no 
$f$-dependence (white noise). 
Fig. 2 (center) shows $S_I$ as a function of $n_s$ at $f=3.16\times10^{-4}$~Hz
and $T=0.13~K$.  At high $n_s$, $S_I$ does not depend on $n_s$ within
the scatter of data.  For $n_s<n_g$, however, an exponential rise of $S_I$ by 
six orders of magnitude is observed with decreasing $n_s$.  This striking 
increase of the slow dynamic contribution to $\sigma$ is consistent 
with the behavior of $\delta\sigma/\langle\sigma\rangle$ (Fig. 2, left).  In 
fact, since $(\delta\sigma)^2/\langle\sigma\rangle^2 = \int S_I(f)df$, it is 
clear that the observed 
enormous increase of the relative rms as $n_s$ is reduced below $n_g$ (Fig. 2,
left) reflects a dramatic slowing down of the electron dynamics.  This is
attributed to the freezing of the electron glass.

Fig. 2 (right) shows another striking feature of our data.  The exponent 
$\alpha$, which describes the frequency dependence of the noise power 
$S_I\propto 1/f^{\alpha}$, 
exhibits a sudden jump at $n_s \approx n_g$.  While
$\alpha\approx 1$ for $n_s>n_g$, $\alpha\approx 1.8$ for $n_s<n_g$; similar
large exponents $\alpha$ have been observed in metallic spin glasses [8].  In general, it
is possible to obtain such high values of $\alpha$ if telegraph noise is
superimposed on flicker ($1/f$) noise~[9].  However, our data exhibit many 
abrupt jumps rather than two-level switching (see Fig. 1, left).  

For $n_s<n_g$, a rapid {\em decrease} of $S_I$ is also observed with 
increasing $T$ (inset to Fig. 2, center), consistent with other studies on Si 
MOSFETs~[10] at higher $T$ ($T=1.5, 4.2$~K).  The observed 
$T$-dependence of noise disagrees with the models of thermally activated 
charge trapping-detrapping fluctuations~[9,11,12], noise generated by 
temperature fluctuations~[13], and in the vicinity of the Anderson 
transition~[14], all of which predict an {\em increase} of noise with 
increasing $T$.  Models of noise in the hopping regime~[15] predict either a 
$T$-independent noise or a power-law decrease of noise with $T$.  In our 
experiment, however, $S_I$ follows an exponential rather than power-law
$T$-dependence.  On the other hand, the observed $S_I (T)$ does agree with the
behavior found in mesoscopic spin glasses~[16,8,17].

\section{Summary}

Measurements of the time dependence of $\sigma$ in low-mobility Si MOSFETs
have shown that, with decreasing $n_s$, a sudden, enormous 
increase in the low-frequency conductivity noise and a sudden shift of the 
spectral weight towards lower frequencies occur at a well-defined density 
$n_g$.  In addition, a dramatic increase of noise with decreasing $T$ have 
been observed for $n_s<n_g$.  Similar behavior in metallic spin glasses
was attributed to spin glass freezing [16,8,17].  For $n_s< n_g$, 
we have also observed long relaxation times and history dependent behavior 
characteristic of a glassy phase, but these effects will be described in 
detail elsewhere. These observations are interpreted as signatures of the
freezing of the electron glass at $n_s=n_g$.  The metal-insulator transition 
occurs at the density $n_c < n_g$ indicating the existence of a metallic 
electron glass, as predicted in~[18].

The authors are grateful to IBM, Yorktown Heights for sample fabrication, and
to V. Dobrosavljevi\'{c} and J. Jaroszy\'{n}ski for useful discussions.  This 
work was supported by NSF Grant DMR-0071668 and by an NHMFL In-House Research
Program grant.

\newpage

\section{Figure captions}

Figure 1

Left: Relative fluctuations of $\sigma$ {\it vs.} time for 
$n_s=3.64\times10^{11}$cm$^{-2}$ at several different $T$ (lowest $T$ at 
bottom, highest $T$ at top).  Different traces have been shifted for clarity.
Right: $<\sigma>$ {\it vs.} $T$ for different $n_s$.  The data for 
many other $n_s$ have been omitted for clarity.  $n_{s}^{\ast}$ and $n_g$ are
marked by arrows.  They were determined as explained in the main text.

Figure 2

Left: Relative rms $\delta\sigma/<\sigma>$ {\it vs.} $n_s$ at 
different $T$.  $n_c$, $n_g$, and $n_{s}^{\ast}$ are marked by arrows.  
Center:  The normalized noise power $S_I$ at $3.16\times 10^{-4}$~Hz {\it vs.}
$n_s$ at $T=0.13$~K; inset: $S_I$ {\it vs.} $T$ for three different 
$n_s (10^{11}$cm$^{-2})$.  Right: The exponent $\alpha$ ($S_I\propto 
1/f^{\alpha}$) {\it vs.} $n_s$.


\begin{thebibliography}{99}\parsep 0pt\parskip -0pt \itemsep 2pt
\bibitem{1}  See E. Abrahams, S. V. Kravchenko, and M. P. Sarachik, 
Rev. Mod. Phys. {\bf 73}, 251 (2001), and references therein.
\bibitem{2}  J. H. Davies, P. A. Lee, and T. M. Rice, Phys. Rev. Lett. {\bf 
49}, 758 (1982); see also M. Pollak and A. Hunt in {\em Hopping Transport in 
Solids}, edited by M. Pollak and B. I. Shklovskii (Elsevier, Amsterdam, 1991),
and references therein.
\bibitem{3}  S. Chakravarty {\it et al.},
Phil. Mag. B~{\bf 79}, 859 (1999).
\bibitem{4} A. A. Pastor and V. Dobrosavljevi\'{c}, Phys. Rev. 
Lett.~{\bf 83}, 4642 (1999).
\bibitem{5} See M. B. Weissman, Rev. Mod. Phys. {\bf 65}, 829 (1993).
\bibitem{6}  S. Bogdanovich and D. Popovi\'{c}, submitted to Phys. Rev.
Lett.; cond-mat/0106545.
\bibitem{7} F. N. Hooge, Phys. Lett. A {\bf 29}, 139 (1969).
\bibitem{8} J. Jaroszy\'{n}ski {\it et al.},
Phys. Rev. Lett. {\bf 80}, 5635 (1998).  
\bibitem{9} See M. B. Weissman, Rev. Mod. Phys. {\bf 60}, 537 (1988).
\bibitem{10}  C. J. Adkins and R. H. Koch, J. Phys. C~{\bf 15}, 1829 
(1982).
\bibitem{11}  See P. Dutta and P. M. Horn, Rev. Mod. Phys. {\bf 53}, 
497 (1981).
\bibitem{12} C. T. Rogers and R. A. Buhrman, Phys. Rev. Lett. {\bf 
53}, 1272 (1984).
\bibitem{13} R. F. Voss and J. Clarke, Phys. Rev. B {\bf 13}, 556 
(1976).
\bibitem{14}  O. Cohen {\it et al.},
Phys. Rev. Lett. {\bf 69}, 3555 (1992); O. Cohen and Z. Ovadyahu, Phys. Rev. B
{\bf 50}, 10442 (1994).
\bibitem{15}  B. I. Shklovskii, Solid State Comm. {\bf 33}, 273 (1980);
V. I. Kozub, Solid State Comm. {\bf 97}, 843 (1996); Sh. Kogan, Phys. Rev. B
{\bf 57}, 9736 (1998).
\bibitem{16}   N. E. Israeloff {\it et al.}, Phys. Rev. Lett. {\bf 63},
794 (1989).
\bibitem{17}  G. Neuttiens {\it et al.}, Phys. Rev. B {\bf 62}, 3905
(2000).
\bibitem{18} V. Dobrosavljevi\'{c} and A. A. Pastor, this volume.
\end{thebibliography}
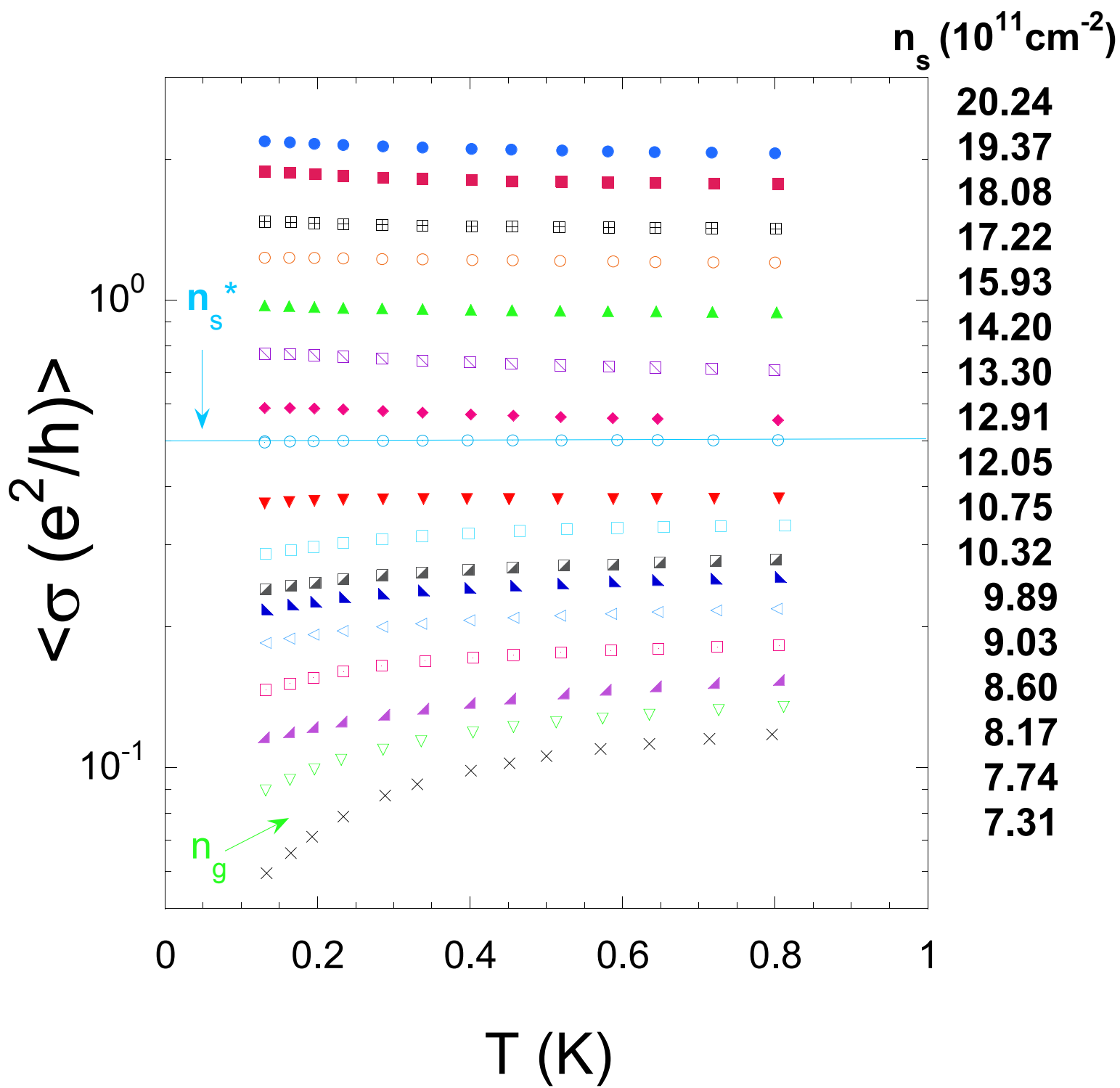
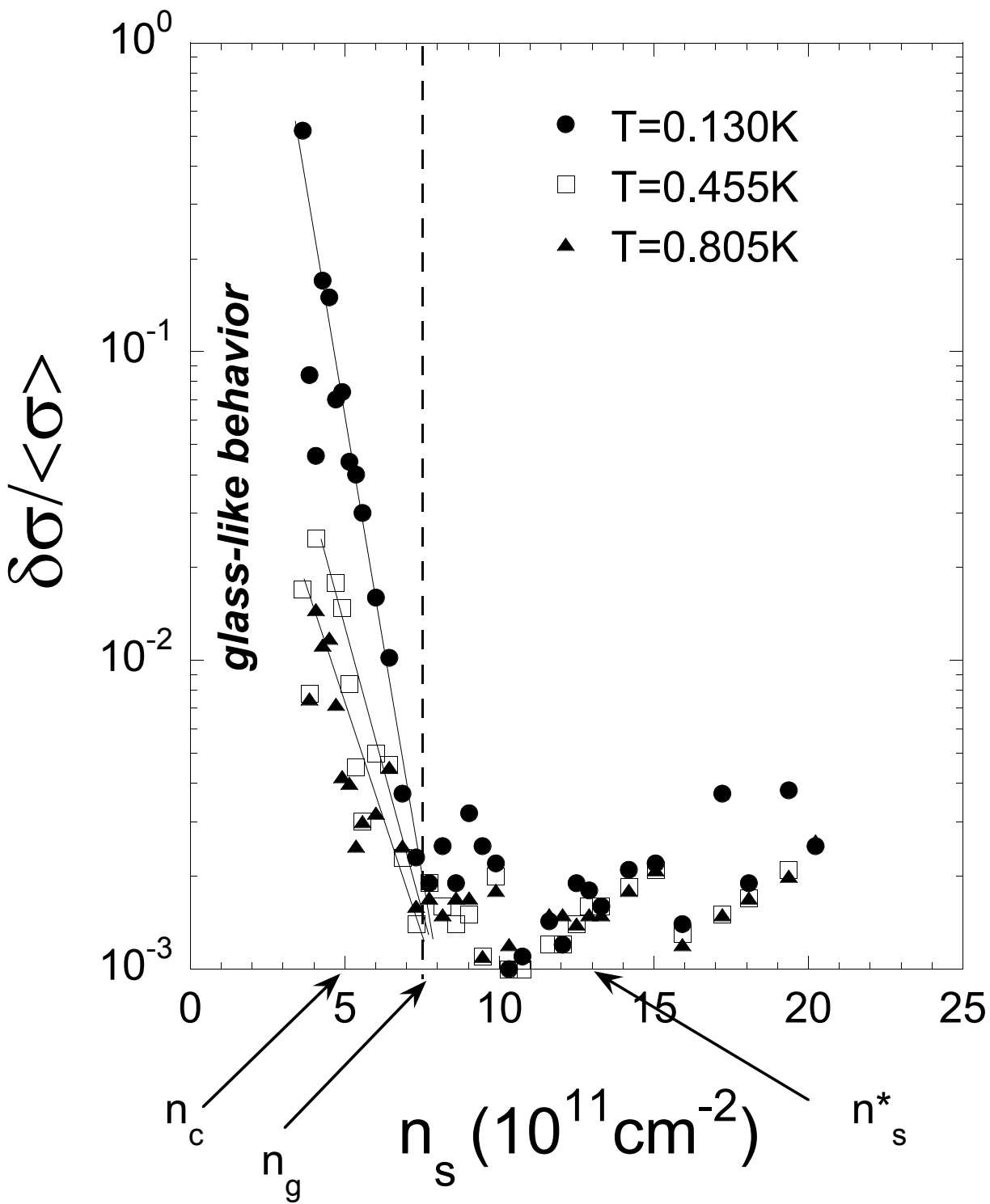
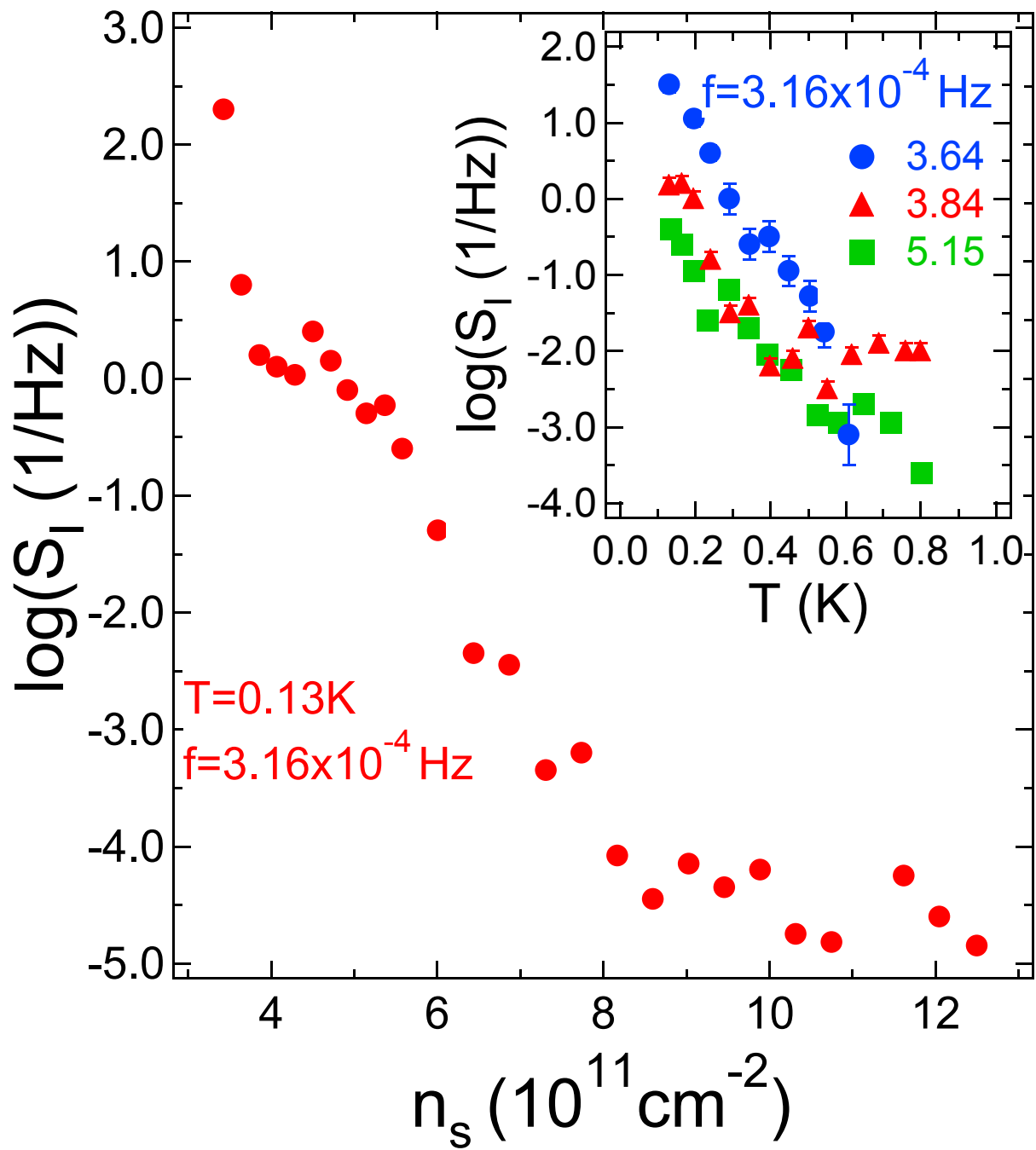
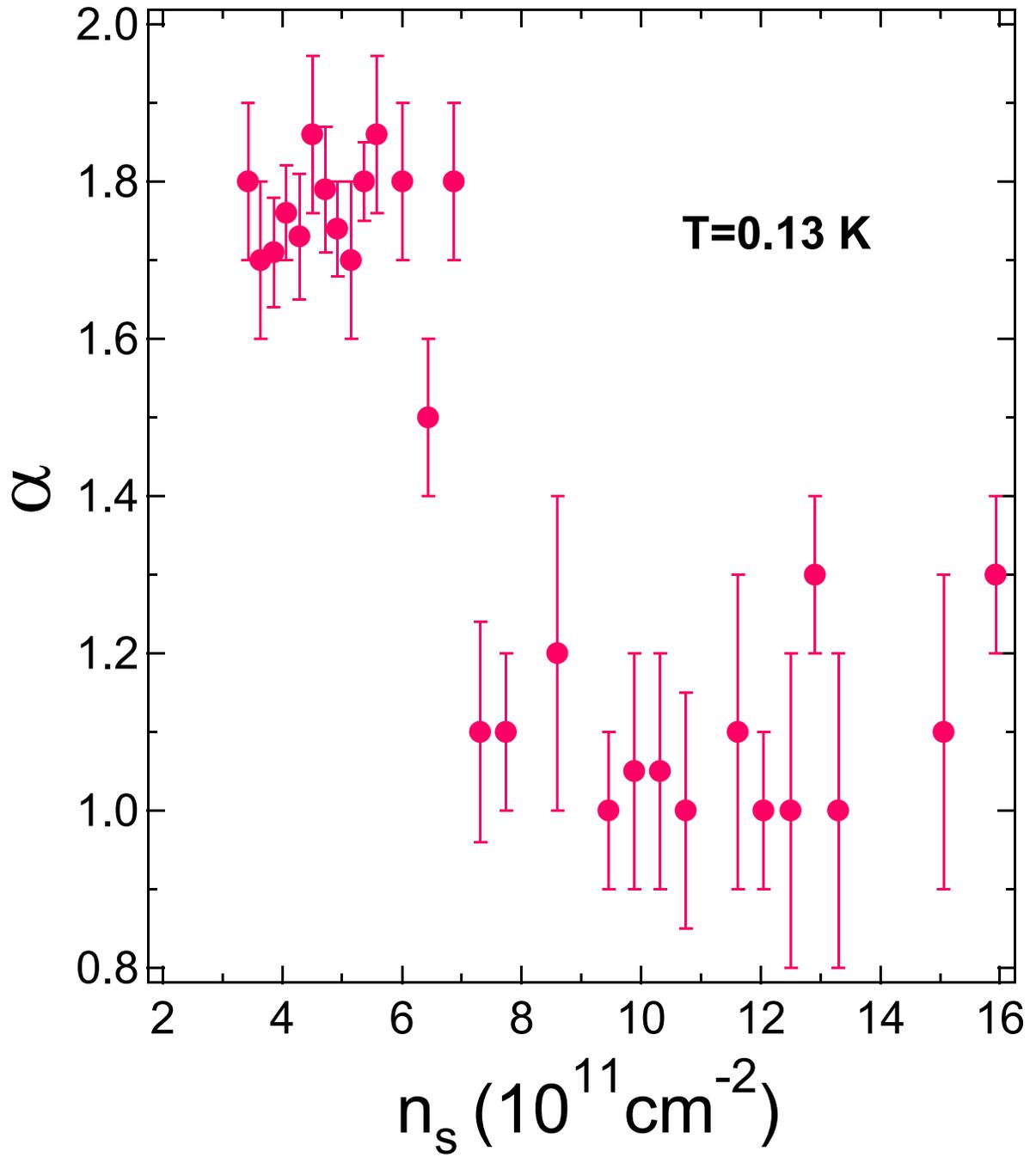
\end{document}